\documentclass[final,5p,times, sort&compress, fleqn,twocolumn]{elsarticle}
\journal{Journal of Magnetism and Magnetic Materials}
\usepackage{amsmath}
\usepackage{bm}
\usepackage{hyperref}
\hypersetup{colorlinks = true}
\hypersetup{breaklinks=true} 
\usepackage{breakurl}

\begin{document}

\begin{frontmatter}

    \title{First-principles study of enhancement of perpendicular magnetic anisotropy
        obtained by inserting an ultrathin LiF layer at an Fe/MgO interface\tnoteref{grant}}

    \author{Yukie Kitaoka}
    \ead{yukie.kitaoka@aist.go.jp}
    \author{Hiroshi Imamura}
    \ead{h-imamura@aist.go.jp}
    \address{National Institute of Advanced Industrial Science and Technology (AIST), Tsukuba, Ibaraki 305-8568, Japan}

    \begin{abstract}
        Perpendicular magnetic anisotropy (PMA) is a key property of magnetoresistive random access memory (MRAM). To increase areal density of MRAM it is important to find a way to enhance the PMA. Recently a strong enhancement of the PMA by inserting an ultrathin LiF layer at an Fe/MgO interface was reported [T. Nozaki et al., NPG Asia Materials (2022) 14: 5]. To understand the origin of the observed enhancement of the PMA we perform first-principles calculations of magetocrystalline anisotropy energy (MAE) of the following four kind of multilayer structures: Fe/MgO, Fe/LiF/MgO, Fe/FeO/MgO, and Fe/FeF/LiF/MgO. We find that the MAEs of the Fe/LiF/MgO and the Fe/FeF/LiF/MgO structures are almost the same as that of the Fe/MgO structure, while the MAE of the Fe/FeO/MgO structure is less than a half of that of the Fe/MgO structure. The results show that the major origin of the enhancement of the PMA obtained by inserting an ultrathin LiF layer at an Fe/MgO interface is the suppression of the mixing of Fe and O atoms at the interface.  We also find that the in-plane Fe-F coupling gives a positive contribution to the MAE while the in-plane Fe-O coupling gives a negative contribution. The results are useful for designing of high-PMA materials.
    \end{abstract}


    \tnotetext[grant]{
        This work was partly based on results obtained from a project, JPNP16007,commissioned by the New Energy and Industrial Technology DevelopmentOrganization (NEDO), Japan.}

\end{frontmatter}
\section{Introduction}
Magnetoresistive random access memory (MRAM) is an emerging nonvolatile memory with low latency, high endurance, low power consumption, and high scalability \cite{Yuasa2013,Ando2014,Kent2015,Apalkov2016,Sbiaa2017,Cai2017,Nozaki2019,Garzon2021,Na2021,Worledge2022}. A MgO-based magnetic tunnel junction (MTJ) is a fundamental element of MRAM because of its large magnetoresistance ratio \cite{Parkin2004,yuasa2004,Ikeda2008} and perpendicular magnetic anisotropy (PMA) \cite{Nakayama2008,Ikeda2010,Meng2011}. In the MgO-based MTJ an insulating MgO layer is sandwiched by two perpendicularly-magnetized ferromagnetic layers. Information is stored as the direction of the magnetization in one ferromagnetic layer called the recording layer. The magnetization of the other ferromagnetic layer called the reference layer is fixed. The stored information is read by using tunnel magnetoresistance (TMR) effect \cite{Julliere1975,Maekawa1982} where the resistance of the MTJ strongly depends on the relative magnetization angle between the two ferromagnetic layers.

In MRAM the MTJs are patterned into nano-pillars. Much effort has been devoted to decreasing the size of the nano-pillar to meet the growing demand of high density integration. Thermal stability of data stored in MRAM is determined by competition between anisotropy energy of the recording layer and thermal energy. The ratio between these two energies is called thermal stability factor defined as $\Delta = K_{\rm PMA} V / k_{\rm B}T$, where $K_{\rm PMA}$ is the effective PMA energy density, $V$ is the volume of the recording layer, $k_{\rm B}$ is the Boltzmann constant, and $T$ is the temperature. In practical application, $\Delta > 60$ is required to guarantee the retention time over 10 years. To satisfy both requirements of high integration density and long retention time it is necessary to design the MTJ with large $K_{\rm PMA}$.

A number of extensive studies have been made for the purpose of increasing the PMA of MTJs both experimentally \cite{Sato2012,Kubota2012,Nozaki2017} and theoretically \cite{Xiang2018,Masuda2018,Su2020,Vojacek2021}. Relevant works before 2017 are  reviewed in Ref. \cite{Dieny2017}. Large PMA was observed in the double barrier structure \cite{Sato2012,Kubota2012}, epitaxial Fe/MgAl$_{\rm 2}$O$_{\rm 4}$ heterostructures \cite{Xiang2018}, and Ir-doped Fe/MgO MTJ \cite{Nozaki2017}. Based on the first-principles calculations several structures were predicted to have large PMA, e.g. the Fe/ MgAl$_{2}$O$_{4}$ interface with W insertion layer \cite{Masuda2018}, MgO-based MTJ with Co/Fe composite layers \cite{Vojacek2021}, and Fe/MgO interface with Pt insertion layer \cite{Su2020}.

Recently Nozaki et al. proposed a novel approach to enhance the PMA based on the atomic-scale interface engineering using fluoride \cite{Nozaki2022}. They fabricated a fully epitaxial Fe(001)/LiF(001)/MgO(001)/Fe(001) structure where a coherent spin-dependent tunneling process was maintained at the LiF/MgO bilayer tunnel barrier. They found that the interface PMA was about 1.4 times higher than that of the standard Fe/MgO interface. Sakamoto et al. performed x-ray magnetic circular dichroism measurements on Fe/LiF/MgO multilayers and found that the LiF insertion increases the orbital-magnetic-moment anisotropy \cite{Sakamoto2022}. However, the origin of the PMA enhancement by insertion of the LiF thin-film at the Fe/MgO interface is still unclear. To understand the origin of the PMA enhancement theoretical analysis based on first-principles calculations is necessary.

\begin{figure}[t]
    \centerline{
        \includegraphics[width=0.9\columnwidth]{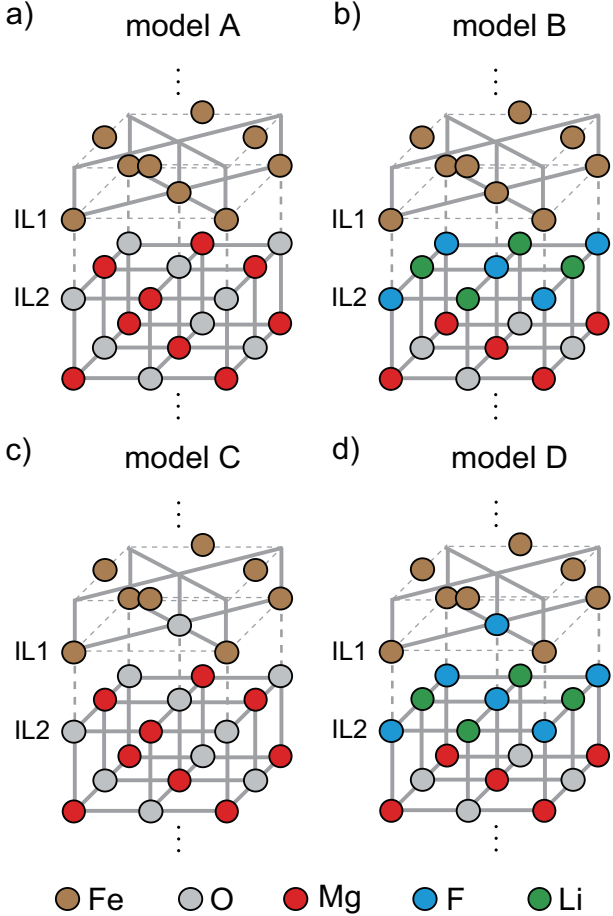}
    }
    \caption{
        Schematic illustrations of crystal structures. Brown, gray, red, green, and blue circles represent the Fe, O, Mg, F, and Li atoms, respectively. Atomic layers at the upper and lower interface are denoted by IL1 and IL2, respectively.
        (a) Model A is the Fe/MgO bi-layer structure where the IL1 is the Fe layer and the
        IL2 is the MgO layer. (b) Model B is the Fe/LiF/MgO tri-layer structure where the IL1 is the Fe layer and the IL2 is the LiF layer. (c) Model C is the Fe/FeO/MgO tri-layer structure where the IL1 is the FeO layer and the
        IL2 is the MgO layer. (d) Model D is the Fe/FeF/LiF/MgO tetra-layer structure where the IL1 is the FeF layer and the IL2 is the LiF layer.
    }
    \label{fig:fig1}
\end{figure}

In this paper, we analyze magnetocrystalline anisotropy energy (MAE) of the four different multilayer structures, Fe/MgO, Fe/LiF/MgO, Fe/FeO/MgO, and Fe/FeF/LiF/MgO, based on first-principles calculations. The calculated MAEs of the Fe/LiF/MgO and the Fe/FeF/LiF/MgO structures are almost the same as the MAE of the Fe/MgO structure, while the MAE of the Fe/FeO/MgO structure is less than a half of the MAE of the Fe/MgO structure. This can be explained by the difference in MAE induced by  the Fe-F and Fe-O in-plane couplings. The results show that the major origin of the PMA enhancement reported in Ref. \cite{Nozaki2017} is the suppression of the mixing of Fe and O atoms at the interface.

\section{Models and Methods}
The crystal structures we analyzed are schematically shown in Figs. \ref{fig:fig1}(a) -- \ref{fig:fig1}(d). The brown, gray, red, green, and blue circles represent the Fe, O, Mg, F, and Li atoms, respectively. All crystal structures consist of eleven atomic layers. Two interface atomic layers, IL1 and IL2, are sandwiched by five atomic layers of the Fe and four atomic layers of the MgO. The model A is the Fe/MgO bi-layer structure where the IL1 is the Fe layer and the IL2 is the MgO layer as shown in Fig. \ref{fig:fig1}(a). The model B is the Fe/LiF/MgO tri-layer structure where the IL1 is the Fe layer and the IL2 is the LiF layer as shown in Fig. \ref{fig:fig1}(b). The model C is the Fe/FeO/MgO tri-layer structure where the IL1 is the FeO layer and the IL2 is the MgO layer as shown in Fig. \ref{fig:fig1}(c). The model D is the Fe/FeF/LiF/MgO tetra-layer structure where the IL1 is the FeF layer and the IL2 is the LiF layer as shown in Fig. \ref{fig:fig1}(d). The in-plane lattice constant is assumed to be the experimentally obtained value of 2.86 \AA\  in bulk Fe. The atomic positions in the out-of-plane direction are fully optimized by using the atomic force calculations until the forces are less than 0.05 eV/ \AA.

Calculations are performed using the full-potential linearized augmented plane-wave (FLAPW) method \cite{Wimmer1981,Li1991,Nakamura2003,Nakamura2010} based on the generalized gradient approximation \cite{Perdew1996} and the scalar relativistic approximation (SRA). The LAPW basis with a cutoff of $\mid$ ${\bf k}$ + ${\bf G}$ $\mid$ $\leq$ 3.9 a.u. and muffin-tin sphere radii of 2.2 a.u. for the Fe atom, 1.40 a.u. for the O and F atoms, and 2.0 a.u. for the Mg and Li atoms are used, where the angular-momentum expansion inside the MT spheres is truncated at $l$ = 8 for the wave functions, charge and spin densities, and potentials. To calculate MAE, the second variational method \cite{Li1990} for treating the spin-orbit coupling (SOC) is performed using the calculated eigenvectors in the SRA. The MAE is determined by the force theorem \cite{Daalderop1990,Wang1996}, which is defined as the energy eigenvector difference for the magnetization oriented along the in-plane [100] and out-of-plane [001] directions.

\section{Results}

The calculated results of MAE for the models A (Fe/MgO), B (Fe/LiF/MgO), C (Fe/FeO/MgO), and D (Fe/FeF/LiF/MgO) are listed in Table \ref{table1}. The MAE in other units is also shown in Table \ref{table3} in \ref{sec:ap} for clarity.
The MAE of the model A (Fe/MgO) is 1.015 meV/unit cell. The MAE of the model B is 0.992 meV/unit cell, which is almost the same as that of the model A. These results suggest that the experimentally observed enhancement of the PMA cannot be  reproduced by the PMA at the Fe/LiF/MgO interface. The MAE of the model C is as small as 0.405 meV/unit cell which is less than a half of the MAE of the Fe/MgO structure. The PMA is strongly suppressed by the oxidation of the Fe layer at the interface as pointed out by Nakamura et al. \cite{Nakamura2010}, and independently by Yang et al. \cite{Yang2011}. On the other hand, the MAE of the model D is 0.979 meV/unit cell which is almost the same as that of the models A and B. The substitution of the Fe atom by the F atom at the interface hardly reduces the PMA.
The results suggest the following scenario for the enhancement of the PMA reported in Ref. \cite{Nozaki2022}. Without insertion of a LiF layer some Fe atoms at the Fe/MgO interface must be substituted by O atoms, which results in the reduction of the PMA. Insertion of a LiF layer forbids the reduction of the PMA due to the oxidation of the Fe layer at the interface. Although some Fe atoms are substituted by F atoms at the interface, the substitution does not reduce the PMA. As a result the PMA of the MgO-based MTJ is enhanced by inserting an ultrathin LiF layer at an Fe/MgO interface.

The calculated results of the orbital-magnetic-moment anisotropy (OMA) of the Fe atoms at the IL1 of the unit cell for the models A, B, C, and D are listed in Table \ref{table1}. The OMAs at the IL1 of the models A, B, C, and D are 0.028, 0.029, 0.021, and 0.021, respectively.
The OMA at the IL1 of the models A and B are larger than that of the models C and D. The results suggest that the insertion of a LiF layer forbids the reduction of OMA due to the oxidation of the Fe layer at the interface.

    {\tabcolsep = 0.4em
        \begin{table}
            \caption{
                \label{table1}
                Calculated results of the magnetocrystalline anisotropy energy (MAE)
                and the orbital-magnetic-moment anisotropy (OMA) of the Fe atoms at the IL1 of the unit cell for the models A, B, C, and D. For clarity, the MAE and OMA in other units are also shown in Table \ref{table3} in \ref{sec:ap}.
            }
            \begin{tabular}{ccccc} \hline \hline
                \rule[-3pt]{0pt}{12pt}
                Model                         & A     & B     & C     & D     \\ \hline
                \rule[-3pt]{0pt}{12pt}
                IL1                           & Fe    & Fe    & FeO   & FeF   \\  
                \rule[-3pt]{0pt}{12pt}
                IL2                           & MgO   & LiF   & MgO   & LiF   \\  \hline
                \rule[-3pt]{0pt}{12pt}
                MAE (meV/unit cell)           & 1.015 & 0.992 & 0.405 & 0.979 \\
                OMA ($\mu_{\rm B}$/interface) & 0.028 & 0.029 & 0.021 & 0.021 \\
                \hline \hline
            \end{tabular}
        \end{table}
    }

\begin{figure}
    \begin{center}
        \includegraphics[width=\columnwidth]{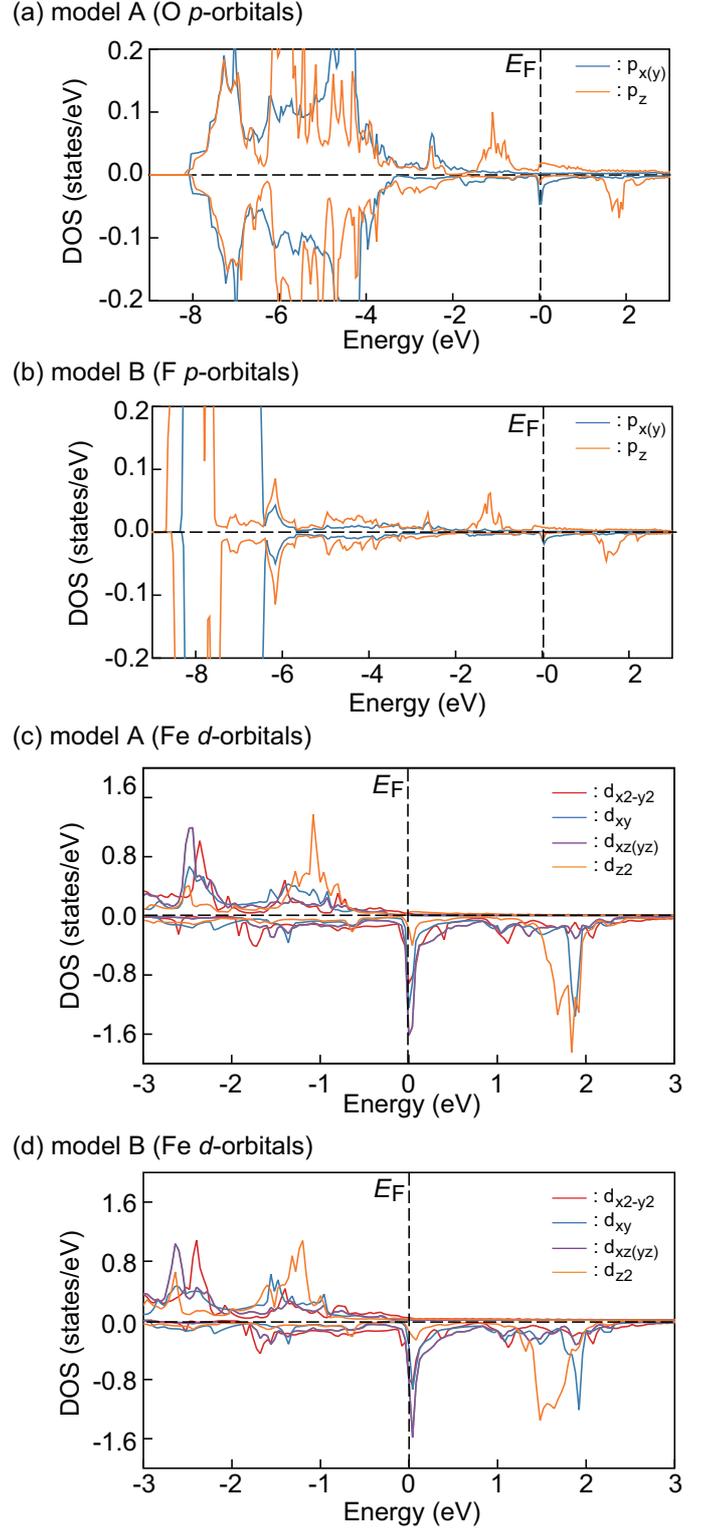}
        \caption{
            Calculated results of the orbital-resolved partial density of states (PDOS) for the models A and B.
            (a) The PDOS of the $p$-orbitals of the O atom in the IL2 of the model A.
            (b) The PDOS of the $p$-orbitals of the F atom in the IL2 of the model B.
            In Panels (a) and (b), the blue and orange curves represent the results of the $p_{x(y)}$- and $p_{z}$-orbitals, respectively.
            (c) The PDOS of the $d$-orbitals of the Fe atom in the IL1 of the model A.
            (d) The PDOS of the $d$-orbitals of the Fe atom in the IL1 of the model B.
            In Panels (c) and (d), the red, blue, purple, and orange curves represent
            the results of the $d_{x^2-y^2}$, $d_{xy}$, $d_{xz(yz)}$, and
            $d_{z^2}$-orbitals, respectively.
            In all Panels, the Fermi energy ($E_{\rm F}$) is set to 0 eV and is indicated by the dashed line.
        }
        \label{fig:fig2}
    \end{center}
\end{figure}

To justify the above scenario it is necessary to understand the physics behind the similarities and differences among the MAE of the models A, B, C, and D listed in Table \ref{table1}. First we discuss the origin of the similarity of the MAE of the models A and B. Figures \ref{fig:fig2}(a) and \ref{fig:fig2}(b) show the calculated partial density of states (PDOS) of the $p$-orbitals of the O and F atoms in the IL2 of the models A and B, respectively. The blue and orange curves represent the results of the $p_{x(y)}$- and $p_{z}$-orbitals, respectively. The Fermi energy ($E_{\rm F}$) is set to 0 eV and is indicated by the dashed line. As shown in Figs. \ref{fig:fig2}(a) and \ref{fig:fig2}(b) the shape of the PDOS around the $E_{\rm F}$ is quite similar to each other while the shape of the PDOS below -2 eV is quite different from each other.
As reported in Ref. \cite{Sakamoto2022}, little spin-magnetic-moment is induced in the O and F atoms. The spin-magnetic-moments in the O and F atoms are 0.026 and 0.023 $\mu_{\rm B}$, respectively, where $\mu_{\rm B}$ is the Bohr magneton.

Figures \ref{fig:fig2}(c) and \ref{fig:fig2}(d) show the PDOS of the $d$-orbitals of the Fe atoms in the IL1 of the models A and B. The red, blue, purple, and orange curves represent the results of the $d_{x^2-y^2}$, $d_{xy}$, $d_{xz(yz)}$, and $d_{z^2}$-orbitals, respectively.
In both models, the PDOS of the $d$-orbitals around the $E_{\rm F}$ concentrates in the minority-spin state and shows a peak in the vicinity of the $E_{\rm F}$. The order of the height of the peaks is the same for both models, i.e. $d_{xz(yz)}$ $>$ $d_{xy}$ $>$ $d_{x^2-y^2}$ $>$ $d_{z^2}$. According to the second-order perturbation theory of MAE, the coupling between the occupied and unoccupied states in the vicinity of $E_{\rm F}$ gives a crucial contribution to MAE \cite{Brooks1940,Bruno1989,Miura2013}. The similarity of PDOS in the vicinity of the $E_{\rm F}$ is the origin of the similarity of the MAE between the models A and B.

    {\tabcolsep = 0.4em
        \begin{table}[b]
            \caption{
                \label{table2}
                Calculated results of the magnetocrystalline energy (MAE) of the Fe, FeO, and FeF mono-atomic layers, which correspond to the IL1 of models A, C, and D, respectively.
            }
            \begin{tabular}{cccc} \hline \hline
                \rule[-3pt]{0pt}{12pt}
                mono-atomic layer   & Fe    & FeO    & FeF   \\ \hline
                MAE (meV/unit cell) & 1.389 & -1.576 & 3.514 \\
                \hline \hline
            \end{tabular}
        \end{table}
    }

Next we discuss the difference of MAE among the models A, C, and D. As shown in Table \ref{table1} the MAE of the model C is less than a half of the MAE of the other models.
The MAE is determined by the arrangement of $d$-orbitals around the $E_{\rm F}$. In the Fe/MgO system, it is already known that the PMA is occurred by the hybridization between Fe $d_{z^2}$-orbital and O $p_{\rm z}$-orbital of the out-of-plane coupling at the Fe-O interface \cite{Shimabukuro2010}. As shown in Figs. \ref{fig:fig2}(c) and \ref{fig:fig2}(d) the arrangements of $d$-orbitals around the $E_{\rm F}$ caused by the out-of-plane coupling of Fe-O (or F) are almost the same. Therefore, we infer that the differences in MAE between the models A, C, and D are related to the modulation of $d$-orbitals by the in-plane coupling. It is necessary to investigate the MAE due to the in-plane coupling.
The models A, C, and D have different atomic compositions in the IL1 layer. The IL1 of the models A, C, and D are the Fe, FeO, and FeF mono-atomic layers, respectively. To concentrate on the effect of the atomic composition on MAE we calculate the MAE of the Fe, FeO, and FeF mono-atomic layers, which correspond to the IL1 of the models A, C, and D, respectively. The results are summarized in Table \ref{table2}. The MAE of the Fe and FeF mono-atomic layers are positive. However, the MAE of the FeO mono-atomic layer is negative, i.e. the FeO mono-atomic layer is in-plane magnetized. The results imply that when the Fe layer at the Fe/MgO interface is oxidized the in-plane Fe-O coupling can reduce the MAE significantly.

\begin{figure}
    \begin{center}
        \includegraphics[width=\columnwidth]{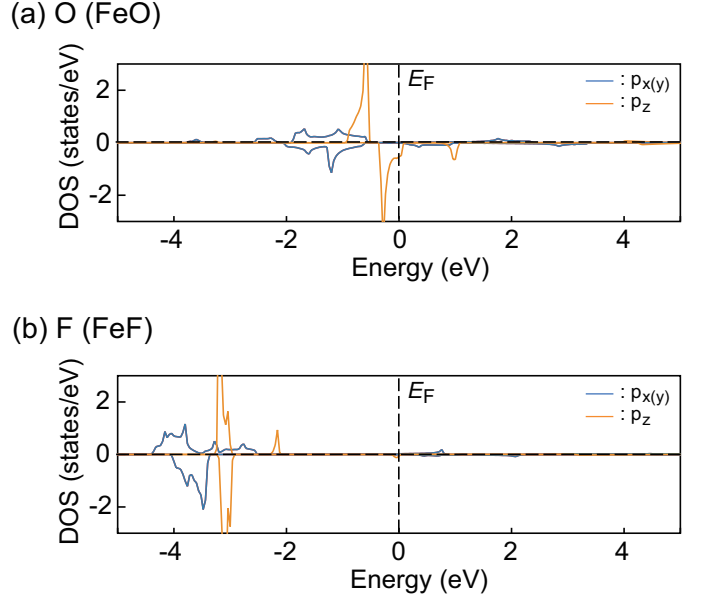}
        \caption{
            Calculated results of the orbital-resolved partial density of states (PDOS) for the FeO and FeF mono-atomic layers. The PDOS of the $p$-orbitals of the O atom in the FeO mono-atomic layer and that of the F atom in the FeF mono-atomic layer are shown in Panels (a) and (b), respectively. The blue and orange curves represent the results of the $p_{x(y)}$- and $p_{z}$-orbitals, respectively. The Fermi energy ($E_{\rm F}$) is set to 0 eV and is indicated by the dashed line.
        }
        \label{fig:fig3}
    \end{center}
\end{figure}

The PDOS of the $p$-orbitals of the O atom in the FeO mono-atomic layer and that of the F atom in the FeF mono-atomic layer are shown in Figs. \ref{fig:fig3}(a) and \ref{fig:fig3}(b), respectively.
The blue and orange curves represent the results of the $p_{x(y)}$- and $p_{z}$-orbitals, respectively. As shown in Fig. \ref{fig:fig3}(a), the $p_{z}$-orbital of the O atom in the FeO mono-atomic layer has the large PDOS at the $E_{\rm F}$, whereas the $p$-orbitals of the F atom in the FeF mono-atomic layer are fully occupied. The absence of the PDOS at the $E_{\rm F}$ implies that the F atom is expected to have little influence on the $d$-orbitals of the Fe atom around the $E_{\rm F}$ and therefore on the MAE.

\begin{figure}
    \begin{center}
        \includegraphics[width=\columnwidth]{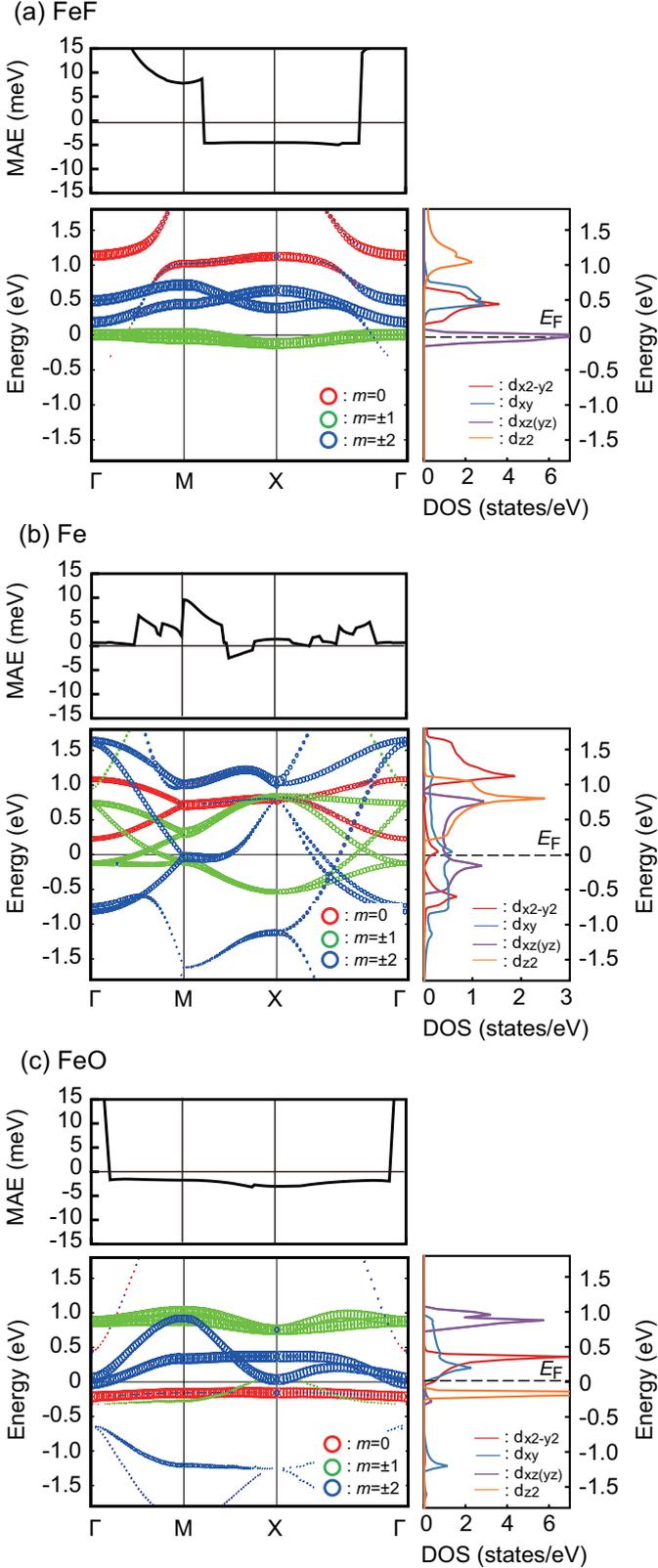}
        \caption{
            Calculated results of the minority-spin band structure of the $d$-orbitals of the Fe atoms, the minority-spin orbital-resolved partial density of states (PDOS) of the $d$-orbitals of the Fe atoms, and the magnetocrystalline anisotropy energy (MAE) contribution along high-symmetry directions for (a) the FeF mono-atomic layer, (b) the Fe mono-atomic layer, and (c) the FeO mono-atomic layer. In the band structure, the red, green, and blue circles represent the results of the magnetic quantum number $m$=0, $\pm$1, and $\pm$2, respectively. In the PDOS, the red, blue, purple, and orange curves represent the results of the $d_{x^2-y^2}$-, $d_{xy}$-, $d_{xz(yz)}$-, and $d_{z^2}$-orbitals, respectively. In all Panels, the Fermi energy ($E_{\rm F}$) is set to 0 eV.
        }
        \label{fig:fig4}
    \end{center}
\end{figure}

The minority-spin band structure and the PDOS of the $d$-orbital of the Fe atoms, and the MAE contribution along the high-symmetry directions for the FeF, Fe, and FeO mono-atomic layers are shown in Figs. \ref{fig:fig4}(a), \ref{fig:fig4}(b), and \ref{fig:fig4}(c), respectively. The panels are arranged from top to bottom in order of decreasing MAE (FeF $>$ Fe $>$ FeO).
In the minority-spin band structure, the red, green, and blue circles represent the results of the magnetic quantum number $m$=0, $\pm$1, and $\pm$2, respectively. In the PDOS, the red, blue, purple, and orange curves represent the results of the $d_{x^2-y^2}$-, $d_{xy}$-, $d_{xz(yz)}$-, and $d_{z^2}$-orbitals, respectively.
For these three models, the band dispersion of $d$-orbitals around the $E_{\rm F}$ is clearly different from each other, which causes the difference in the MAE. As pointed out by Wang et al. \cite{Wang1993}, the SOC interaction between the occupied and unoccupied $d$-states for the same spin-state with the same (different) magnetic quantum number gives a positive (negative) contribution to the MAE.
For the FeF mono-atomic layer, the states with $m$=$\pm$1 ($d_{xz(yz)}$-orbitals) are localized at the $E_{\rm F}$ as shown in Fig. \ref{fig:fig4}(a). A large positive MAE, i.e. PMA, is obtained by the SOC interaction between the occupied and unoccupied minority-spin states with the same magnetic quantum number of $m$=$\pm$1, which is located around $\Gamma$ and M. The result implies that fluoridation of the Fe layer at the Fe/LiF interface will not reduce the MAE.
For the Fe mono-atomic layer, the PDOS around the $E_{\rm F}$ shows broad peaks for all $d$-orbitals due to the Fe-Fe hybridization as shown in Fig. \ref{fig:fig4}(b).
The positive and negative contributions from the SOC interaction give the positive MAE as a total.
On the other hand, the PDOS of the FeO mono-atomic layer shows that very little $d$-orbitals in the minority-spin state crosses the $E_{\rm F}$ as shown in Fig. \ref{fig:fig4}(c). As a result, the positive contribution to the MAE from the SOC interaction between the occupied and unoccupied states with the same magnetic quantum number, $m$=$\pm$2 ($d_{z^2}$-orbital), exists only in a small region around $\Gamma$. At the other k-points, the SOC interactions between the different magnetic quantum numbers of the $m$=0 ($d_{z^2}$-orbital) and $m$=$\pm$1 ($d_{xz(yz)}$-orbitals) gives the negative contributions to the MAE.
Hence, the MAE of the MTJ is reduced by oxidation of the Fe layer at the interface because of the negative contribution to MAE from the in-plane Fe-O coupling.

\section{Summary}
In summary, we investigate the origin of the enhancement of the PMA by insertion of an ultrathin LiF layer at an Fe/MgO interface which was reported in Ref. \cite{Nozaki2022}. Using the first-principles calculations based on the FLAPW method and force theorem we calculate the MAE of the following four kind of multilayer structures: Fe/MgO, Fe/LiF/MgO, Fe/FeO/MgO, and Fe/FeF/LiF/MgO. We show that the MAE of the Fe/LiF/MgO and the Fe/FeF/LiF/MgO structures is almost the same as that of the Fe/MgO structure, while the MAE of the Fe/FeO/MgO structure is less than a half of that of the Fe/MgO structure. To clarify the contributions of the in-plane Fe-O and Fe-F couplings to the MAE we also calculate the MAE of three kind of mono-atomic layers: Fe, FeO, and FeF. We find that the in-plane Fe-F coupling gives a positive contribution to the MAE while the in-plane Fe-O coupling gives a negative contribution. The results show that the major origin of the enhancement of the PMA obtained by inserting an ultrathin LiF layer at an Fe/MgO interface is the suppression of the mixing of Fe and O atoms at the interface.

\section*{Acknowledgements}
The authors thank T. Nozaki and T. Nozaki of AIST for discussions on their experimental results.

\appendix
\setcounter{table}{0}

\section{Table of MAE and OMAin other units}
\label{sec:ap}
In Table \ref{table3}, the calculated results of MAE listed in Table \ref{table1} are presented in three different units and the calculated results of OMA listed in Table \ref{table1} are presented in two different units for clarity.

    {\tabcolsep = 0.4em
        \begin{table}[h]

            \caption{
                \label{table3}
                Calculated results of the magnetocrystalline anisotropy energy (MAE) and the orbital-magnetic-moment anisotropy (OMA) of the Fe atoms at the IL1 of the unit cell of the models A, B, C, and D. For clarity, the results for MAE are presented in three different units and the results for OMA are presented in two different units. Note that the number of the Fe atoms at the IL1 of the unit cell of the models A and B is twice that of models C and D.
            }
            \begin{tabular}{ccccc} \hline \hline
                \rule[-3pt]{0pt}{12pt}
                Model                         & A      & B      & C      & D      \\ \hline
                \rule[-3pt]{0pt}{12pt}
                IL1                           & Fe     & Fe     & FeO    & FeF    \\  
                \rule[-3pt]{0pt}{12pt}
                IL2                           & MgO    & LiF    & MgO    & LiF    \\  \hline
                \rule[-3pt]{0pt}{12pt}
                MAE (meV/unit cell)           & 1.015  & 0.992  & 0.405  & 0.979  \\
                MAE (meV/Fe atom)             & 0.0846 & 0.0827 & 0.0368 & 0.0890 \\
                MAE (mJ/m$^{\rm 2}$)          & 1.988  & 1.943  & 0.794  & 1.917  \\
                OMA ($\mu_{\rm B}$/Fe atom)   & 0.014  & 0.015  & 0.021  & 0.021  \\
                OMA ($\mu_{\rm B}$/interface) & 0.028  & 0.029  & 0.021  & 0.021  \\
                \hline \hline
            \end{tabular}
        \end{table}
    }



\end{document}